\documentclass[12pt]{article}

\usepackage{epsfig} 
\usepackage{amsbsy} 

\def\beq{\begin{equation}}
\def\eeq{\end{equation}}
\def\eq{\beq\eeq}
\def\beqn{\begin{eqnarray}}
\def\eeqn{\end{eqnarray}}

\def\({\left(}
\def\){\right)}

\def\ba{\begin{eqnarray}}
\def\ea{\end{eqnarray}}
\def\bq{\begin{equation}}
\def\eq{\end{equation}}
\def\lsim{\mathrel{\raisebox{-.6ex}{$\stackrel{\textstyle<}{\sim}$}}}
\def\sla#1{\ifmmode%
\setbox0=\hbox{$#1$}%
\setbox1=\hbox to\wd0{\hss$/$\hss}\else%
\setbox0=\hbox{#1}%
\setbox1=\hbox to\wd0{\hss/\hss}\fi%
#1\hskip-\wd0\box1 }

\def\asb{{}\ifmmode \bar{\alpha}_s \else $\bar{\alpha}_s$\fi}
\def \as   {\ifmmode \alpha_s \else $\alpha_s$ \fi}

\def\so3#1{\,{\rm S}_{1,\,3}\left(#1 \right)}
\def\st2#1{\,{\rm S}_{2,\,2}\left(#1 \right)}

\newskip\humongous \humongous=0pt plus 1000pt minus 1000pt

\newif\ifdtup

\catcode`\@=11

\@addtoreset{equation}{section}

\def\theequation{\thesection.\arabic{equation}}

\def\@normalsize{\@setsize\normalsize{15pt}\xiipt\@xiipt
\abovedisplayskip 14pt plus3pt minus3pt%
\belowdisplayskip \abovedisplayskip
\abovedisplayshortskip \z@ plus3pt%
\belowdisplayshortskip 7pt plus3.5pt minus0pt}

\def\small{\@setsize\small{13.6pt}\xipt\@xipt
\abovedisplayskip 13pt plus3pt minus3pt%
\belowdisplayskip \abovedisplayskip
\abovedisplayshortskip \z@ plus3pt%
\belowdisplayshortskip 7pt plus3.5pt minus0pt
\def\@listi{\parsep 4.5pt plus 2pt minus 1pt
     \itemsep \parsep
     \topsep 9pt plus 3pt minus 3pt}}

\@twosidetrue

\catcode`@=12

\catcode`\@=11

\def\section{\@startsection{section}{1}{\z@}{3.5ex plus 1ex minus
   .2ex}{2.3ex plus .2ex}{\large\bf}}

\def\thesection{\arabic{section}}
\def\thesubsection{\arabic{section}.\arabic{subsection}}
\def\thesubsubsection{\arabic{section}.\arabic{subsection}.\arabic{subsubsection}}

\def\appendix{\setcounter{section}{0}
 \def\thesection{\Alph{section}}
 \def\theequation{\Alph{section}.\arabic{equation}}
 \def\thesubsection{\Alph{section}.\arabic{subsection}}
\def\thesubsubsection{\Alph{section}.\arabic{subsection}.\arabic{subsubsection}}

\def\section{\@startsection{section}{1}{\z@}{3.5ex plus 1ex minus
   .2ex}{2.3ex plus .2ex}{\large\bf}}
}

\newcommand{\ccaption}[2]{
  \begin{center}
    \parbox{0.85\textwidth}{
      \caption[#1]{\small\it {#2}}}
  \end{center}    }

\def \to   {\mbox{$\rightarrow$}}

\newcount\minutes
\newcount\scratch

\def\timestamp{%
\scratch=\time
\divide\scratch by 60
\edef\hours{\the\scratch}
\multiply\scratch by 60
\minutes=\time
\advance\minutes by -\scratch
---$\,$\hours:\null
\ifnum\minutes< 10 0\fi
\the\minutes}

\begin{document}
\begin{titlepage}
\nopagebreak
{\flushright{
        \begin{minipage}{5cm}
         MADPH 01-1226 \\
	 BNL-HET-01/15\\
         MSUHEP-10510\\	
         DFTT 13/2001\\
        {\tt hep-ph/0105129}\hfill \\
        \end{minipage}        }

}
\vfill
\begin{center}
{\LARGE \bf \sc
 \baselineskip 0.9cm
$H+2$~jets via gluon fusion
          
}
\vskip 0.5cm 
{\large  
V.~Del Duca$^a$, W.~Kilgore$^b$, C.~Oleari$^c$, C.~Schmidt$^d$ and 
D.~Zeppenfeld$^c$
}  
\vskip .2cm 
{$^{(a)}$ {\it I.N.F.N., Sezione di Torino
via P.~Giuria, 1 - 10125 Torino, Italy}}\\ 
{$^{(b)}$ {\it Physics Department,
  Brookhaven National Laboratory, 
  Upton, New York 11973, U.S.A.}}\\
{$^{(c)}$ {\it Department of Physics, University of Wisconsin, Madison, WI
53706, U.S.A. }}\\  
{$^{(d)}$ {\it Department of Physics and Astronomy,
Michigan State University,
East Lansing, MI 48824, U.S.A.}}\\

\vskip
1.3cm    
\end{center}

\nopagebreak
\begin{abstract}
Real emission corrections to $gg\,\to\, H$, which lead to $H+2$~jet events,
are calculated at order $\alpha_s^4$. Contributions include top-quark 
triangles, boxes and pentagon diagrams and are evaluated analytically for
arbitrary top mass $m_t$. 
This new source of $H+2$~jet events is compared to the weak-boson fusion 
cross section for a range of Higgs boson masses. The heavy 
top-mass approximation appears to work well for intermediate 
Higgs-boson masses, provided that the transverse momenta of the final-state 
partons are smaller than the top-quark mass.
\end{abstract}
\vfill
\vfill
\end{titlepage}
\newpage                                                                     

\section{Introduction}
Gluon fusion via a top-quark loop is expected to be the most copious source 
of Higgs bosons in high energy hadronic collisions. Because of its 
importance for Higgs searches at the CERN LHC~\cite{CMS,ATLAS}, precise 
knowledge of higher order QCD corrections will be crucial to predict
event characteristics for a Higgs signal, and to extract Higgs properties,
like Higgs couplings to gauge bosons and fermions~\cite{Zeppenfeld:2000td}.

For the measurement of Higgs boson couplings, the separation of different
Higgs production processes is necessary. Weak-boson fusion
(WBF), i.e. the radiation of a Higgs boson from a $t$-channel $W$ or
$Z$ boson in electroweak quark-quark scattering, is characterized by the 
presence of two 
forward jets. Such $H+2$~jet events can also arise from ${\cal O}(\alpha_s^2)$
corrections to gluon fusion. For a measurement of WBF cross sections, the 
``background'' from this latter process must be known, i.e. we need to 
calculate the cross section and event characteristics of $H+2$~jet events
arising from gluon fusion. 
 
Next-to-leading order (NLO) QCD corrections to the inclusive gluon-fusion 
cross section are known to 
be large, leading to a $K$-factor close to two~\cite{HggNLO}. Because the
lowest order process is loop induced, a full NNLO calculation would entail
a three-loop evaluation, which presently is not feasible. Fortunately,
for the intermediate Higgs mass range, which is the focus of present interest,
the Higgs boson mass $m_H$ is small compared to the top-quark pair threshold
and the large $m_t$ limit promises to be an adequate approximation. 
Consequently, present efforts on a NNLO calculation of the inclusive 
gluon-fusion cross section
concentrate on the $m_t\to\infty$ limit, in which the task reduces to an 
effective two-loop calculation~\cite{H2loop}. 
In order to assess the validity
of this approximation, gluon-fusion cross-section calculations, which include
all finite $m_t$ corrections, are needed. 
Of particular interest are 
phase space regions where one or several of the kinematical invariants 
are of the order of, or exceed, the top-quark mass, i.e. regions of
large Higgs boson or jet transverse momenta, or regions where dijet invariant 
masses become large. 

In this letter we present first results of such a calculation, namely 
the evaluation of the real emission corrections to
gluon fusion which lead to $H+2$~parton final states, at order 
$\alpha_s^4$.
This includes the subprocesses
\bq
\label{eq:processes}
qQ\,\to \,qQH\;,\qquad qg \,\to \,qgH\;, \qquad gg\,\to \,ggH\;,
\eq
and all crossing-related processes. The corresponding scattering amplitudes,
involving triangle, box and pentagon diagrams with a top-quark loop, are
calculated analytically for arbitrary values of the top-quark mass. 

\begin{figure}[htb]
\centerline{
\epsfig{figure=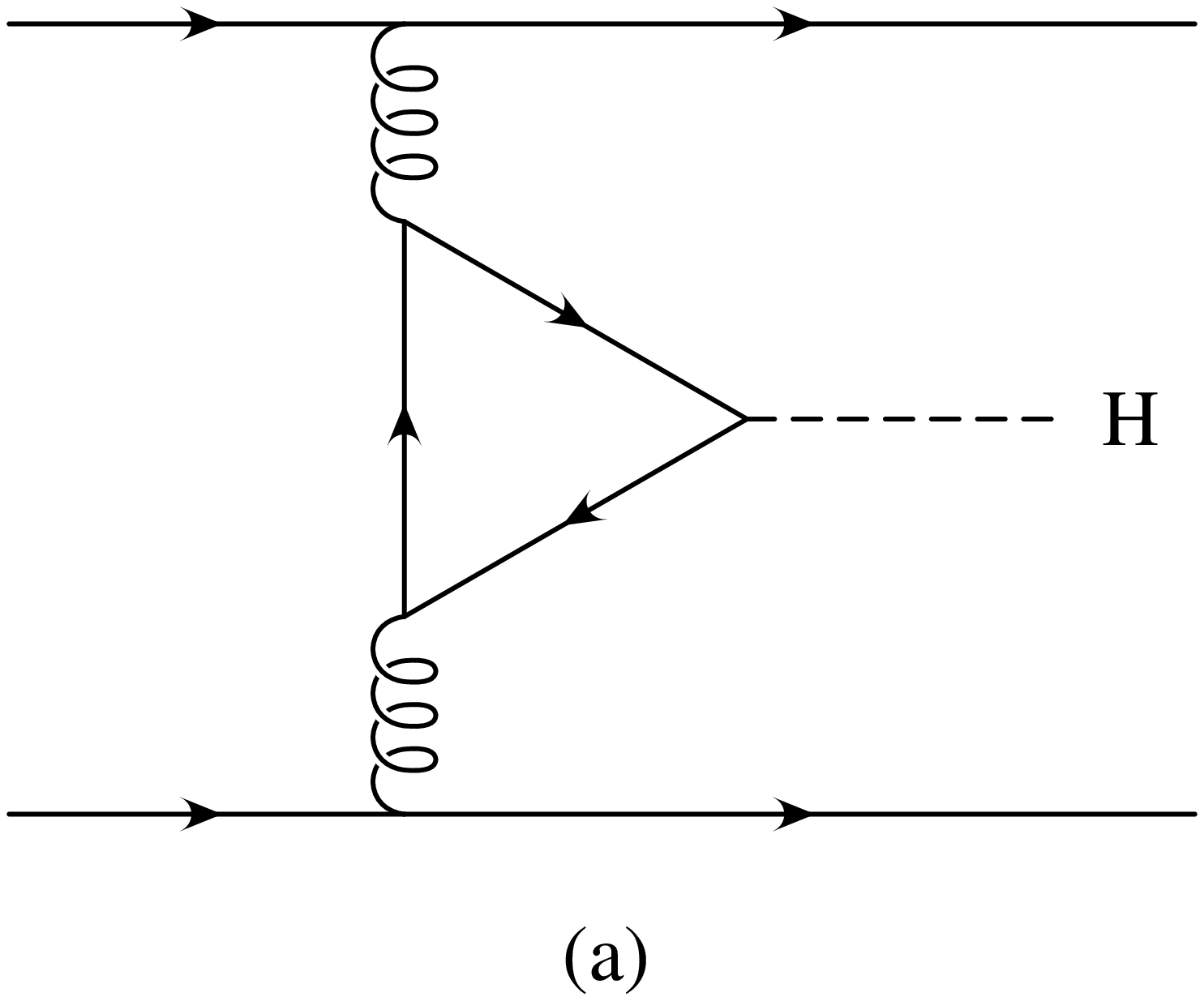,width=0.3\textwidth,clip=} \ \ 
\epsfig{figure=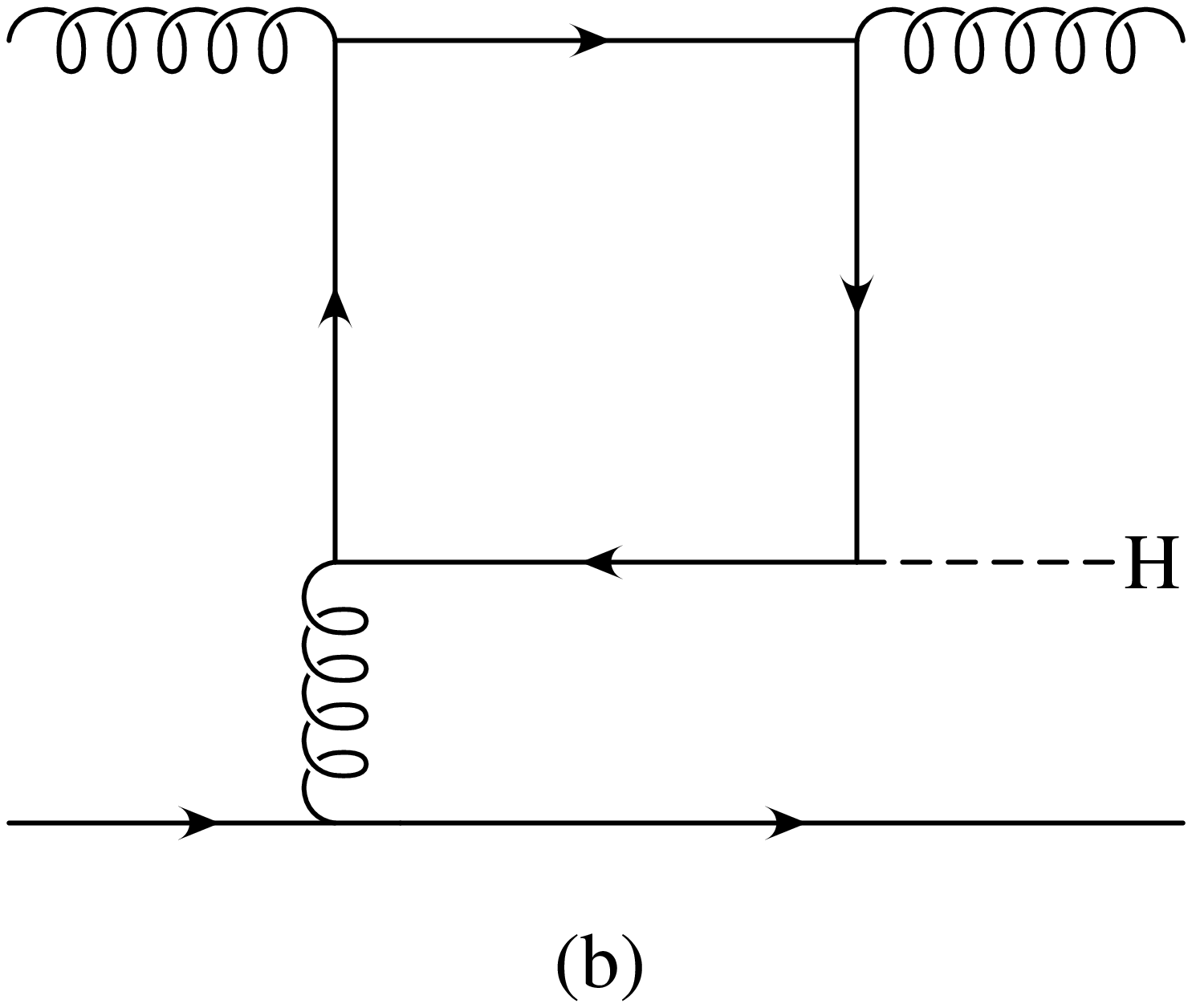,width=0.3\textwidth,clip=} \ \ 
\epsfig{figure=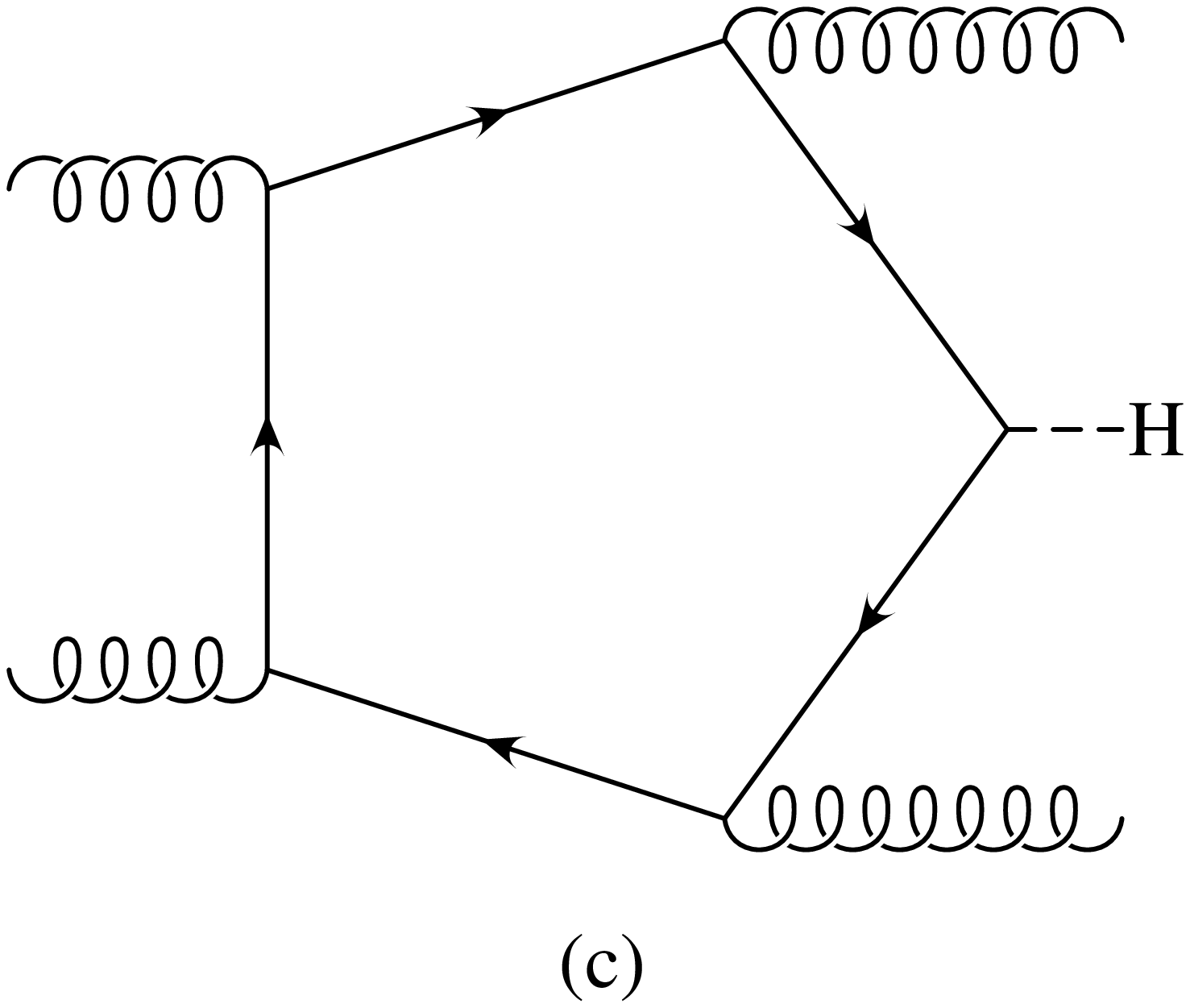,width=0.3\textwidth,clip=} \ \ 
}
\ccaption{}
{ \label{fig:feyn} Samples of Feynman graphs contributing to $H+2$~jet
production via gluon fusion. }
\end{figure}

\section{Calculation}
In Fig.~\ref{fig:feyn} we have collected a few representative Feynman
diagrams that occur in  Higgs production plus two jets, at order
$\as^4$. In our calculation, the top quark is treated as massive but we 
neglect all other quark masses, so that the 
Higgs couples to gluons only via a top-quark loop. Typically we have a
$ggH$ coupling through a triangle loop (Fig.~\ref{fig:feyn}~(a)), a $gggH$
coupling through a box loop (Fig.~\ref{fig:feyn}~(b)) and a $ggggH$ one
through a pentagon loop (Fig.~\ref{fig:feyn}~(c)).

The number and type of diagrams can be easily traced back if one thinks to
insert these Higgs-gluon ``vertices'' into the tree-level diagrams for $2
\, \to \, 2$ QCD parton scattering. \\
In the following counting, we exploit
Furry's theorem, i.e. we are counting as one the two charge-conjugation 
related diagrams where the loop momentum is 
running clockwise and counter-clockwise. This halves the number of diagrams. 
In addition, the crossed processes are not listed as extra diagrams, but are
included in the final results.

\begin{enumerate}
 \item {\boldmath $qQ \,\to\, qQ H$} There is only one diagram obtained from
 the insertion of a triangle loop into the tree-level diagram $ qQ \,\to\, qQ
 $, i.e.\ Fig.~\ref{fig:feyn}~(a).

 \item {\boldmath $qg \,\to\, qg H$} At tree level, there are 3 diagrams
 contributing to the process $qg \,\to\, qg$: one with a three-gluon vertex and
 two Compton-like ones.
 Inserting a triangle loop into every gluon line, we have a total of 7
 different diagrams.
 In addition, we can insert a box loop into the diagram with the
 three-gluon vertex, in 3 different ways: the $3!$ permutations of the 
 3 gluons are reduced to 3 graphs by using Furry's theorem. 
 In total we have 10 different diagrams for the $qg \,\to\, qg H$ contribution.

 \item {\boldmath $gg \,\to\, gg H$} Four diagrams contribute to
 the tree-level scattering process $gg\,\to\, gg$: a four-gluon vertex
 diagram and 3 diagrams with two three-gluon vertices each.
 Inserting a triangle loop in any of the gluonic legs, gives rise to 19
 different diagrams.
 The insertion of the box loop in the 3 diagrams with three-gluon vertices
 yields another 18 diagrams.
 Finally, there are 12 pentagon diagrams (corresponding to $4!$ permutations 
 of the external gluons, divided by 2, according to Furry's theorem).
\end{enumerate}

The procedure for computing these diagrams is outlined below. A detailed 
discussion is postponed to a later paper~\cite{H2jet_future}.\\

{\bf Tensor integrals}
Triangle and box integrals are evaluated in terms of Passarino-Veltman 
$C_{ij}$ and $D_{ij}$ functions~\cite{PV}. 
Gluon polarization vectors are expressed 
in terms of external momenta, which substantially simplifies expressions. 
Helicity techniques are employed for the numerical evaluation of all
amplitudes involving external quarks. 

In case of the pentagon diagrams appearing in the $ggggH$ amplitudes,
the four external gluon-momenta form a 
basis in terms of which we expand the gluon polarization vectors. 
In this way, all the scalar products appearing in the numerator of the tensor
integrals are products of the external momenta and the loop momentum
($k$). These dot products are then written as a difference of two propagators, 
plus a $k$-independent term.  In this way, tensor pentagon integrals are
transformed into a linear combination of tensor boxes and of scalar pentagons.

{\bf Scalar integrals}
All the scalar integrals needed for the calculation 
are finite in $D=4$ dimensions, due to the presence of the top-quark mass.
No further regulator is required.
Scalar triangles and boxes have been known for a long time in the
literature~\cite{tris_boxes} and efficient computational procedures are
available~\cite{Denner}.
Following the procedure outlined in Refs.~\cite{BDK}, we have
expressed all scalar pentagons as linear combinations of scalar boxes.

{\bf Checks} 
We were able to perform two different kinds of checks on the analytic
amplitudes we computed: 
\begin{enumerate}

\item {\em Gauge invariance:} an important test of the calculation is the
gauge 
invariance of the 
amplitudes for $qg\,\to\, qgH $ and $gg\,\to\, ggH $. This test
was performed both analytically and numerically, in the final Fortran
program.

\item {\em Large $m_t$ limit:} the amplitudes for
Higgs plus two partons agree in the large $m_t$ limit with the corresponding
amplitudes obtained from the heavy-top effective Lagrangian~\cite{kauffman}.  
This was done numerically by setting $m_t=3$~TeV.
We found 
agreement within a few
percent, when the Higgs mass is varied in the range 100~GeV $< m_H <$ 700~GeV.
\end{enumerate}

\section{Discussion of the results}
The gluon-fusion processes at ${\cal O}(\alpha_s^4)$, together with 
weak-boson fusion ($qQ\,\to\, qQH$ production via $t$-channel exchange of a $W$
or $Z$), 
are expected to be the dominant sources of $H+2$~jet events at the LHC. Cross
sections for the former diverge as the final-state partons become collinear
with one another or with the incident beam directions, or as final-state
gluons become soft. A minimal set of cuts on the final-state partons, which
anticipates LHC detector capabilities and jet finding algorithms, is required
to define an $H+2$~jet cross section. Our minimal set of cuts is
\bq \label{eq:cuts_min}
p_{Tj}>20\;{\rm GeV}, \qquad |\eta_j|<5,\qquad R_{jj}>0.6,
\eq
where $p_{Tj}$ is the transverse momentum of a final state jet
and $R_{jj}$ describes the separation of the two partons in the 
pseudo-rapidity $\eta$ versus azimuthal angle plane. 

In our simulation, we used the CTEQ4L set for parton-distribution 
functions~\cite{cteq4},
with a factorization scale equal to $\sqrt{p_{T1} \, p_{T2} }$ and 
we fixed $\as=0.12$. We postpone any discussion on factorization- and
renormalization-scale dependence to a further
paper~\cite{H2jet_future}.

\begin{figure}[htb]
\centerline{\epsfig{figure=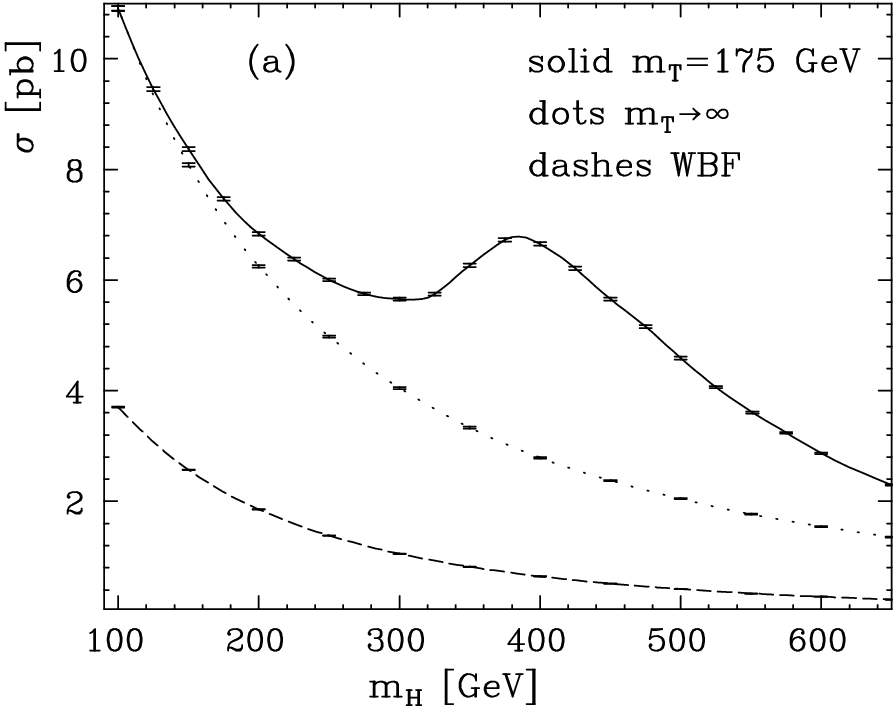,width=0.495\textwidth,clip=}
\epsfig{figure=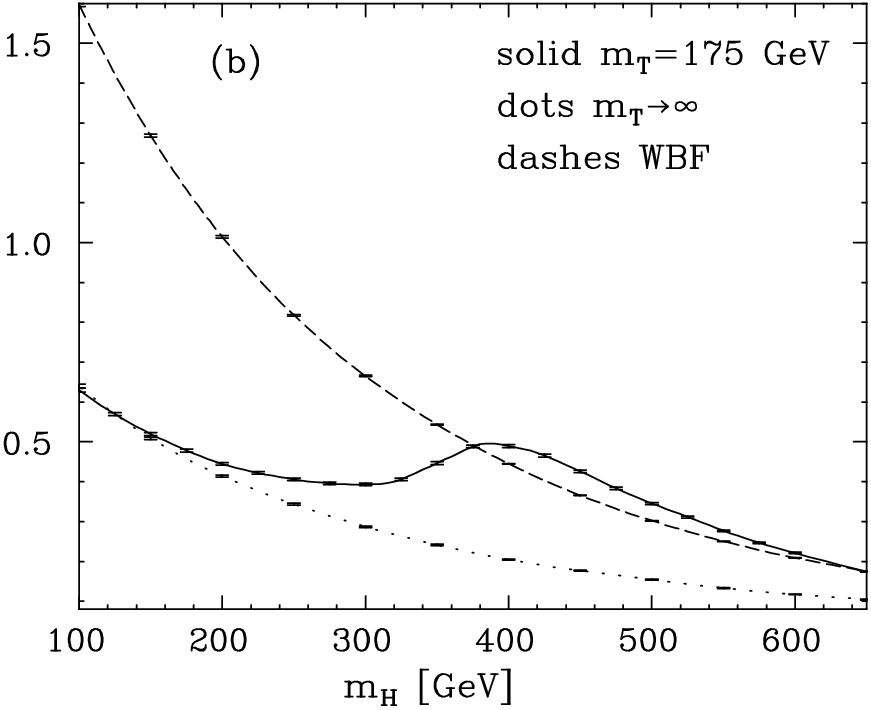,width=0.48\textwidth,clip=}
}
\ccaption{}
{ \label{fig:sigmaMh} 
$H+2$~jet cross sections in pp collisions at 
$\protect\sqrt{s}=14$~TeV as a function of the Higgs boson mass.
Results are shown for gluon-fusion processes induced by a top-quark loop
with $m_t=175$~GeV and in the $m_t\,\to\,\infty$ limit, computed using the
heavy-top effective Lagrangian, 
and for weak-boson fusion. The two panels correspond to two sets of jet cuts:
(a) inclusive selection (see 
Eq.~(\protect\ref{eq:cuts_min})) and (b) WBF selection 
(Eqs.~(\protect\ref{eq:cuts_min}) and (\protect\ref{eq:cut_gap})).}
\end{figure}

Expected $H+2$~jet cross sections at the LHC are shown in
Fig.~\ref{fig:sigmaMh}, as a function of the Higgs boson mass, $m_H$.  The
three curves compare results for the expected SM gluon-fusion cross section
at $m_t=175$~GeV (solid line) with the large $m_t$ limit (dotted line), and
with the WBF cross section (dashed line).
Error bars indicate the statistical errors of the Monte Carlo integration. 
Cross sections correspond to the sum over all Higgs decay modes: finite Higgs
width effects are included.

Figure~\ref{fig:sigmaMh} (a) shows cross sections within the minimal cuts of
Eq.~(\ref{eq:cuts_min}). The gluon-fusion contribution dominates because
the cuts retain events with jets in the central region, 
with relatively small dijet invariant mass. 

In order to assess background 
levels for WBF events it is more appropriate to consider typical tagging 
jet selections employed for WBF studies~\cite{RZ_WW}. This is done in 
Fig.~\ref{fig:sigmaMh} (b) where, in addition to the cuts of 
Eq.~(\ref{eq:cuts_min}), we require
\bq \label{eq:cut_gap}
|\eta_{j1}-\eta_{j2}|>4.2, \qquad \eta_{j1}\cdot\eta_{j2}<0, \qquad
m_{jj}>600\;{\rm GeV},
\eq
i.e. the two tagging jets must be well separated, with three units of 
pseudo-rapidity between the jet definition cones, they must reside in 
opposite detector hemispheres and they must possess a large dijet 
invariant mass. 
With these selection cuts the weak-boson 
fusion processes dominate over gluon fusion by about 3/1 for 
Higgs boson masses in the 100 to 200~GeV range. This means that a relatively
clean separation of weak-boson fusion and gluon-fusion processes will be 
possible at the LHC, in particular when extra central-jet-veto techniques are 
employed to further suppress semi-soft gluon radiation in QCD backgrounds.
A suppression by a factor three  of gluon fusion as compared to WBF
cross sections is to be expected with a central-jet veto~\cite{RZ_WW}.

A conspicuous feature of the $H+2$~jet gluon-fusion cross sections in 
Fig.~\ref{fig:sigmaMh} is the threshold enhancement at $m_H\approx 2m_t$,
an effect which is familiar from the inclusive gluon-fusion cross section.
Near this ``threshold peak'' the gluon-fusion cross section rises to equal
the WBF cross section, even with the selection cuts of Eq.~(\ref{eq:cut_gap}).
Well below this region, the large $m_t$ limit provides an excellent 
approximation to the total $H+2$~jet rate from gluon fusion, at least when 
considering the total Higgs production rate only. Near top-pair threshold
the large $m_t$ limit underestimates the rate by about a factor of 2.

\begin{figure}[htb]
\centerline{\epsfig{figure=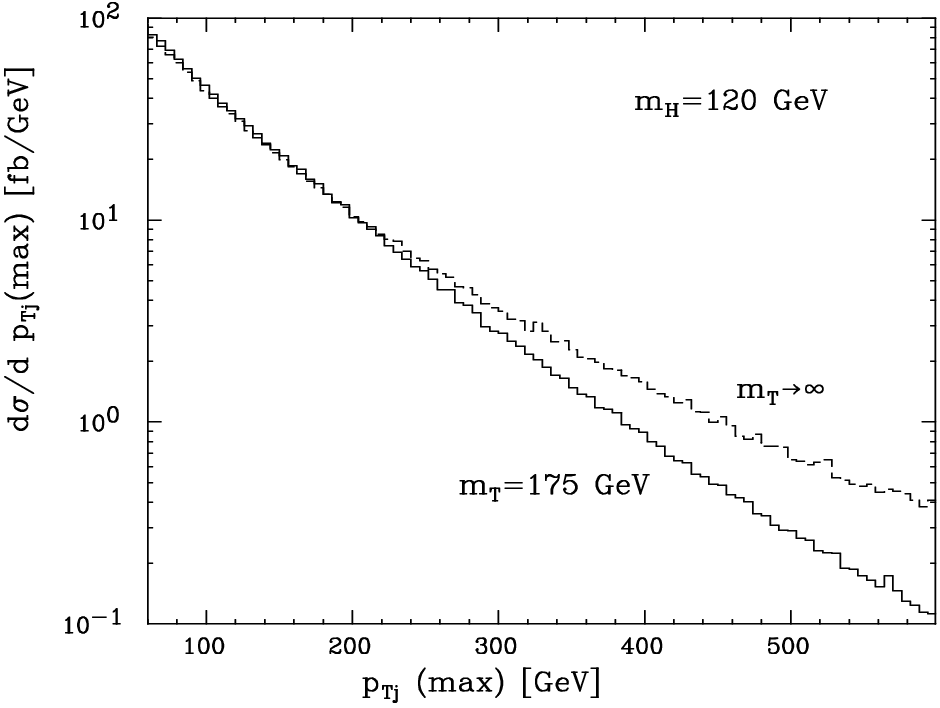,width=0.75\textwidth,clip=}}
\ccaption{} { \label{fig:ptjmax} Transverse-momentum distribution of the
hardest jet in $H+2$~jet events from gluon-fusion processes. Jets are defined
via the inclusive selection of Eq.~(\protect\ref{eq:cuts_min}). The two
curves are for $m_t=175$~GeV and for the $m_t \,\to\,\infty $ limit, 
computed using the
heavy-top effective Lagrangian. The mass of the Higgs is set to 120~GeV.}
\end{figure}

A somewhat surprising feature of Fig.~\ref{fig:sigmaMh} (b) is the excellent
approximation provided by the large $m_t$ limit at Higgs boson masses below
about 200~GeV. Naively one might expect the large dijet invariant mass,
$m_{jj}>600$~GeV, and the concomitant large parton center-of-mass energy 
to spoil the $m_t\,\to\,\infty$ approximation. This is not the case, however. 
The validity range of the $m_t\,\to\,\infty$ approximation is best appreciated 
in Fig.~\ref{fig:ptjmax}, where we show the transverse-momentum distribution
of the harder of the two jets for an intermediate mass Higgs boson.
At transverse momenta below 200~GeV the large $m_t$ limit works extremely
well. It is the large $p_{Tj}$ region where this approximation breaks down.
Small jet transverse momenta but large dijet invariant masses are well
modeled by the  
large $m_t$ limit.

\section{Conclusion}
In this letter we have provided first results of a full ${\cal O}(\alpha_s^4)$
calculation of $H+2$~jet cross sections, including the analytic top-mass
dependence. Event rates are sizable, of order 5 to 10~pb with minimal 
jet-selection cuts. With stringent forward-jet tagging cuts, as suggested 
for the Higgs search in weak-boson fusion, rates drop to about a third of the
WBF rate, which implies that a relatively clean separation of gluon fusion 
and weak-boson fusion will be possible at the LHC. Sizable deviations from
the large $m_t$ limit are seen for Higgs boson masses of order $2m_t$ 
or for large jet transverse momenta. However, the large $m_t$
limit works well for parton center of mass energies $\sqrt{\hat s}>2m_t$,
provided  that $m_H \lsim m_t$ and 
that the transverse momenta remain small compared to $m_t$.

\subsection*{Acknowledgements}
C.S. acknowledges the U.S. National Science Foundation under grant
PHY-0070443.
W.K. acknowledges the DOE funding under Contract 
No.~DE-AC02-98CH10886. 
This research was supported in part by the University of Wisconsin Research
Committee with funds granted by the Wisconsin Alumni Research Foundation and
in part by the U.~S.~Department of Energy under Contract
No.~DE-FG02-95ER40896. 


\end{document}